\documentclass[]{aanew} 
\usepackage{graphicx} 
\usepackage[authoryear]{natbib}
\bibpunct{(}{)}{;}{a}{}{,}
\begin{document} 

\headnote{Research Note}
\title{ Further Wolf-Rayet stars in the starburst cluster 
\object{Westerlund~1} 
\thanks{Based on observations collected at the European Southern 
Observatory, La Silla, Chile (ESO 71.D-0151)}} 

\author{I.~Negueruela\inst{1}
\and
J.~S.~Clark\inst{2,3}} 
 
\institute{ Departamento de F\'{\i}sica, Ingenier\'{\i}a de Sistemas y
  Teor\'{\i}a de la Se\~{n}al, Universidad de Alicante, Apdo. 99,
  E03080 Alicante, Spain 
\and
Department of Physics and Astronomy, The Open University, Walton Hall,
Milton Keynes MK7 6AA, UK
\and
Department of Physics and Astronomy, University College London, 
Gower Street, London, WC1E 6BT, England, UK }

\mail{ignacio@dfists.ua.es}
\titlerunning{More WR stars in Wd~1} 

\date{Received    / Accepted     } 
 
\abstract{We present new low and intermediate-resolution spectroscopic
  observations of the Wolf Rayet (WR) star population  
in the massive starburst cluster Westerlund~1. Finding charts are
  presented for five new WRs -- four WNL and one WCL -- raising  
the current total of known WRs in the cluster to 19. We also present
  new spectra and correct identifications for the majority of the 14
  WR stars previously known, notably confirming the presence of two
  WNVL stars. Finally we briefly discuss the massive star population of
  Westerlund~1 in comparison to other massive young galactic
  clusters.} 

\maketitle 

\section{Introduction} 
 
Wolf-Rayet (WR) stars represent the last stage in the evolution of
 massive stars and are characterised by very high temperatures and
 exhaustive mass loss \citep[see][ for a recent review]{hucht}. In
 spite of  their interest, our current understanding of their
 evolutionary paths is still rather limited. The evolutionary links
 between different WR subtypes are not well established and, more
 importantly, the correspondence between observed characteristics and
 position on theoretical tracks is still unclear \citep[cf.][]{mm03}. 

Observation of WR stars in open clusters and comparison
 with other massive members have yielded most of the constraints on
 which understanding of these objects is based, providing ages and
 progenitor masses \citep[e.g.,][]{mas01}. Unfortunately, the number
 of WR stars in clusters is relatively small and generally each
 cluster contains only one or two WR stars, resulting in rather poor
 evolutionary constraints.  

The young open cluster \object{Westerlund 1} (henceforth Wd~1;
\citealt{west61}) offers the possibility of studying an important
population of WR stars of a given age and chemical
composition within the context of a large homogeneous population of
massive stars. This highly reddened cluster is found at a distance of
between 2 and 5~kpc (most likely close to the upper limit; see
\citealt{main}) and contains a large number of evolved
massive stars.  

\defcitealias{one}{Paper~I}

Spectra of 11 WR stars obtained with the ESO 1.5-m telescope were
presented by Clark \& Negueruela (\citeyear{one}; henceforth
\citetalias{one}). Because of the small size of the 
telescope and the low spatial resolution of the configuration used (as
well as the lack of appropriate finding charts for the field), the
identification of some of these objects was problematic, as many WR
spectra appeared to arise from what were obviously unresolved blends of
stars. Moreover the Signal-to-Noise Ratio (SNR) of several of the spectra
was low, allowing only very approximate spectral classifications. The
objects observed comprised 5 late WC stars (WCL), 5 late WN stars
(WNL) and one broad-lined, presumably early, WN star.

\citet{main} reported the identification of three further objects
with WR-like characteristics. Two of them had been observed
serendipitously at intermediate resolution as they fell on the slit
when observations of brighter objects were taken. Their spectra were
therefore of very low SNR, but were suggestive of a WCL and a
transitional Ofpe object, belonging to a class recently rechristened
as very late WN stars (WNVL). The third object had only been observed
at very low resolution and it appeared as an OB supergiant with very
strong emission lines, also suggestive of a transitional object.

Here we present higher quality spectra of the majority of these
objects, taken under exceptional
seeing conditions, allowing a much better characterisation of the WR
population in \object{Wd~1}. We also present spectra of 5 new WR stars
found in the field of \object{Wd~1}. This brings the total number of
WR stars known in \object{Wd~1} to 17 + two transitional objects. In
what follows, we will adopt the naming convention of \citet{main},
but will drop the word ``candidate'' from the name of 
those Wolf-Rayet stars whose identification has been secured.

\section{Observations \& data reduction} 
 
Observations of \object{Wd~1} were obtained with the ESO Multi-Mode
Instrument (EMMI) on the 3.5-m New Technology Telescope (NTT) at La
Silla, Chile. They were taken during a run on 2003 June 5th-8th,
though only the 6th and the 7th were useful because of cloud cover. On
the night of the 6th, though some high cirrus were present, the seeing
was exceptionally good, staying below 0\farcs6 for most of the night
and reaching $<$0\farcs4 at times. Imaging of the cluster area was
obtained using the $R$ and H$\alpha$ (\#654) filters.

  \begin{figure}
\resizebox{\hsize}{!}{\includegraphics[angle=-90]{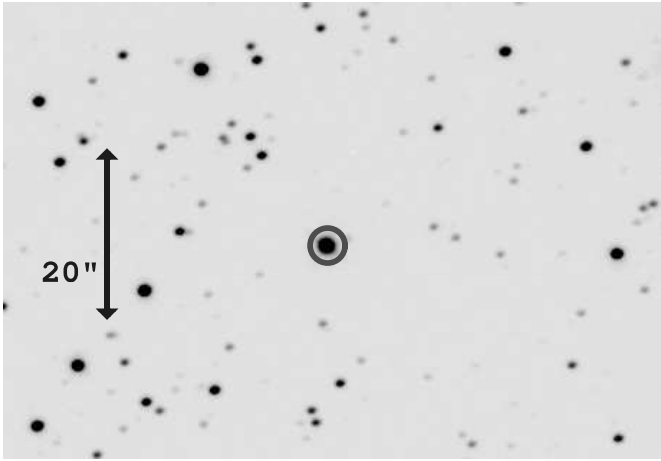}}
\label{fig:south}
\caption{$R$-band finding chart for WR star N, an outlier to the South
of the cluster. This object is catalogued as USNO-B1.0
0440-0523445.
}
\end{figure}

Due to the varied science goals of the observations and the serendipitous 
nature of many of the detections, spectra of a number of sources were 
obtained using a varied set of instrumental 
configurations. For intermediate resolution spectra, we used
the red arm with gratings \#6 and \#7. On the night of
June 6th, grating \#6 covered the $\lambda\lambda6440-7140${\AA}
range. On June 7th,
grating \#6 covered $\lambda\lambda8225-8900${\AA} and 
grating \#7, $\lambda\lambda6310-7835${\AA}.

For low resolution, we used the red arm with grisms \#1 and \#4. Grism
\#1 covers the $\lambda\lambda3850-10000${\AA} range, with a
resolution $R=263$.  Grism
\#4 covers the $\lambda\lambda5550-10000${\AA} range, with a
resolution $R=613$. Note, however, that because of the high reddening,
the signal-to-noise ratio rapidly decreases in the blue end of the
spectra - effectively limiting  the spectra to  
$\lambda > 6500$\AA\ for grism \#4 and $\lambda > 5500$\AA\ for grism 
\#1.

An observation of the field to the South of the cluster was also conducted
in slitless spectroscopy mode. This technique, based on the use of a
low dispersion grism (in our case, \#1) coupled with a broad-band
filter (Bessel $R$) resulting in an ``objective 
prism-like'' spectrogram of all the objects in the field, has been used
by \citet{bp01} to search for emission line stars in open
clusters. 

Image pre-processing was carried out with {\em MIDAS} software, while
data reduction was achieved with the
{\em Starlink} packages {\sc ccdpack} and  \citep{draper} and {\sc figaro}
\citep{shortridge}. Analysis was carried out using {\sc figaro} and {\sc dipso}
\citep{howarth}.

The results of this inevitably somewhat varied observational approach
was the identification of a further 5  
WR stars within Wd~1. Inspection of the slitless image led to the
location of three obvious candidates lying in
the outskirts of the cluster, which were later confirmed as WR stars
by long-slit spectroscopy. A fourth WR star was found serendipitously
while obtaining spectra of objects in the Southern part of the
cluster. Following the notation used in \citetalias{one}, these are WR
stars N, O, P and Q. Finally, as part of a  
dedicated investigation into the apparent Ofpe star W14, we identified
a final WR, designated R. 


\begin{table} 
\begin{center} 
\begin{tabular}{cl} 
\hline 
Object      & Wavelength range (\AA)\\ 
\hline 
WR A & 5550$-$10000\\
WR B & 5550$-$10000\\
WR C & 5550$-$10000\\
WR K & 5550$-$10000\\
& 6440$-$7140\\
WR L & 5550$-$10000\\
WR M & 3850$-$10000\\
&5550$-$10000\\
&6440$-$7140\\
WR N & 3850$-$10000\\
WR O & 3850$-$10000\\
WR P & 3850$-$10000\\
WR Q & 5550$-$10000\\
WR R & 5550$-$10000\\
W5 & 5550$-$10000\\
&6440$-$7140\\
&6310$-$7835\\
\hline 
\end{tabular} 
\caption{Observation log, presenting all the spectroscopic
observations for each target. See text for the configurations resulting in 
each
spectral range.}
\end{center} 
\end{table}

  \begin{figure*}
\resizebox{\hsize}{!}{\includegraphics[angle=0]{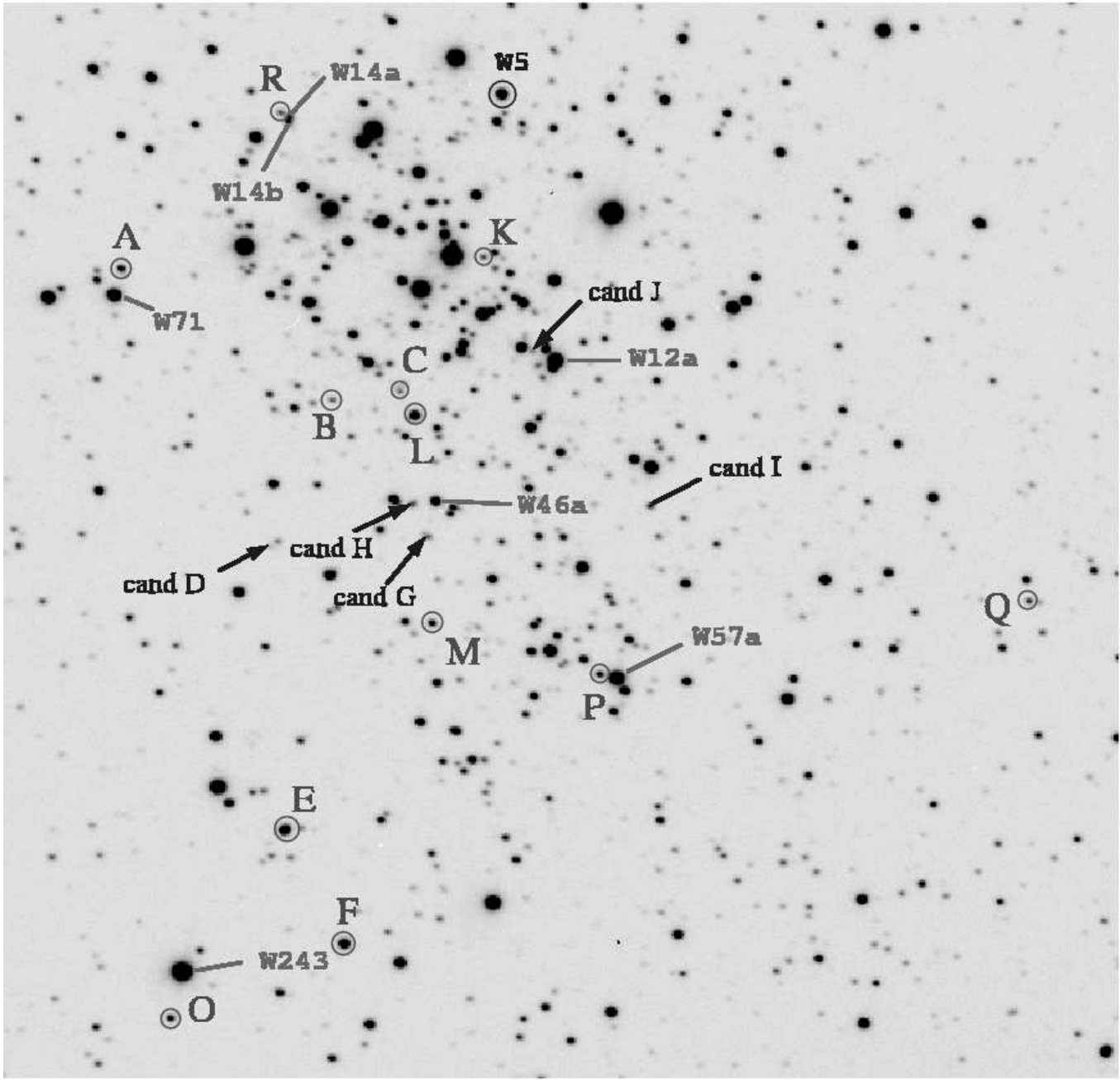}}
\label{fig:cluster}
\caption{$R$-band finding chart for WR stars in the central region of
\object{Wd~1}. Stars surrounded by circles are secure identifications
of the WRs, while objects for which the identification is not secure
(in the sense defined in the text) are marked with the word
``cand''. A few other objects, useful for the discussion, are
indicated, following the naming convention of \citetalias{one}. W5,
which could be a very late WN or an extreme B supergiant is also circled.}
\end{figure*}

For aesthetic reasons, the finder in Fig.~2 is based on an $R$-band image
obtained with VLT UT1/FORS2 on June 10th 2004, in service mode.

\section{Results}
\label{sec:res}

Due to crowding and the low spatial resolution of the Boller \&
Chivens spectrograph, the exact identifications of some of the WR
stars presented in \citetalias{one} and \citet{main} were uncertain. The
WR features were 
observed in spectra attributable to a blend of several objects,
an example of this being the identification of the
continuum of an O-type supergiant and emission lines typical of a WR
star in the  low-resolution spectrum of W14 \citep{main}.

 This situation has been somewhat alleviated with the current dataset. We
now can confirm the exact identification for all but 5 of the
currently identified WR stars. 
For the purposes of this paper, we consider that a WR identification
is sufficiently secure when either several spectra of the same star
exist, all displaying the WR features, or a spectrum with sufficiently
high spatial resolution allows the identification of a single
candidate. This is the case for WR star N (Fig.~1), and
all the objects circled in Fig.~2 (also marked by their
corresponding letters). For those objects without such
secure identification, the word ``cand'' is shown in front of the
corresponding letter, with an arrow pointing to the most likely
identification. 

Of the previously identified WRs, the new observations have confirmed
the positions of WRs 
A, B, C, E, F, L and M, while the most probable candidates for WRs D,
G, I and J remain unchanged. However, we present a new 
identification for WR K and suggest a different candidate for WR H.

Moving to the new WRs, star N is an outlier some $4\farcm5$ South of the
cluster. This object is listed in the USNO catalogue as USNO-B1.0
0440-0523445, with no measured blue magnitudes, red magnitudes $r_{1}=16.6$
and $r_{2}=16.9$ and infrared magnitude $i=13.0$.  These magnitudes and
colour make it a very likely cluster member, further supported by its
late-WC classification (see Section~\ref{sec:wcs}). A finder for this
object is shown in Fig.~1.

Another nearby object, USNO-B1.0 0440-0523458, was found to display
emission lines, but its spectrum shows it to be a foreground Be
star. Its foreground character is confirmed by the relatively low
reddening ($b_{2}=14.08$, $r_{2}=13.04$).

WR stars O and P are within the area of the cluster previously
explored and their positions are identified in Fig.~2; no 
photometry is available for either object. Their spectra are shown in
Fig.~3. 

WR star Q lies on the Southwestern reaches of the cluster and it is one
of the westernmost likely members identified by \citet{main}, who give
the following magnitudes $B=23.7$, $V=20.3$, $R=17.5$, $I=14.7$. Its
spectrum is shown in Fig.~4 and discussed in
Section~\ref{sec:wns}. Its location with respect to the main body of
the cluster can be seen in Fig.~2.

Finally, in our new images, W14 is clearly resolved into three objects of
similar brightness, which we designate W14a, b and c.
(see Fig.~2 for identifications).
Individual spectra of all three objects have been obtained and  WR
features are unmistakably associated with the Easternmost object,
W14c which we designate as WR R; its spectrum is displayed in
Fig.~4, but no photometry is available for the
individual components of W14. 

\subsection{Spectral Classification}

\begin{figure} 
\resizebox{\hsize}{!}{\includegraphics[bb=110 110 540 715, angle=-90]{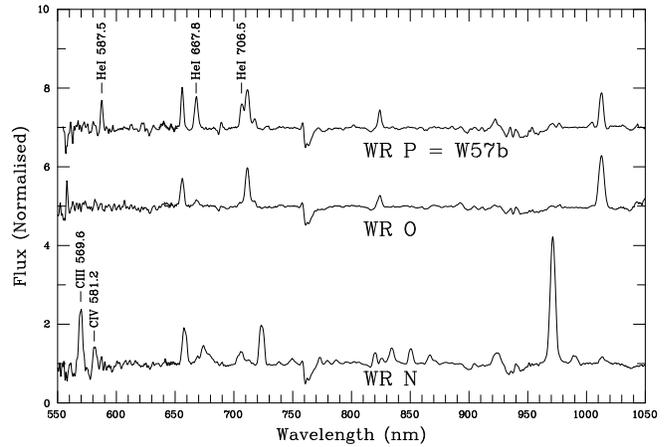}} 
\label{fig:lowres} 
\caption{Spectra of three WR stars that were observed only with grism
  \#1. WR stars P \& O are
  WNs, while WR star N is a late WC. } 
\end{figure} 

Our new spectroscopic observations -- listed in Table~1 -- allow the
classification of the newly identified  
WRs (WRs N-Q), while permitting a more accurate analysis of stars
previously observed at lower resolution   
and/or S/N (WRs A, B, C, K--M).  The results are summarised in Table~2.

\subsubsection{WC stars}
\label{sec:wcs}

\begin{figure*}
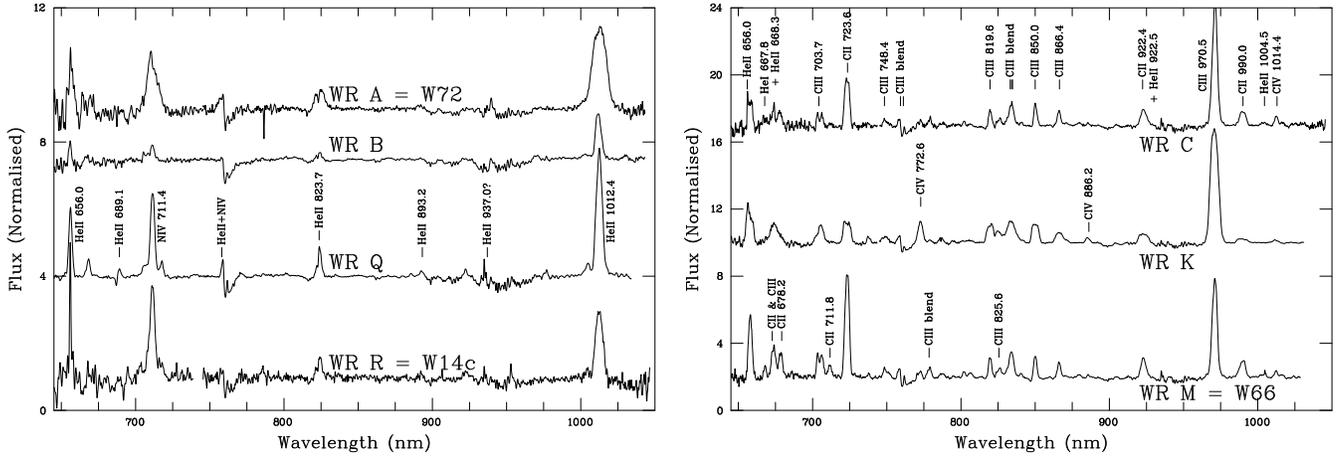
 
\begin{center}
\begin{minipage}[b]{0.5\textwidth}
\resizebox{\hsize}{!}{\includegraphics[bb=110 110 540 715, angle=-90]{fig4a.ps}} 
\end{minipage}%
\begin{minipage}[b]{0.5\textwidth}
\resizebox{\hsize}{!}{\includegraphics[bb=110 110 540 715, angle=-90]{fig4b.ps}}
\end{minipage}
\end{center}
\label{fig:gr4} 
\caption{Grism \#4 spectra of WR stars in Wd~1. {\it Left: WN
  stars}. Stars Q and R are  
  WN6-7, while the spectral types of A and B are not certain. Star A
  is earlier than WN7 and star B could be WN8+OB.
 {\it Right: WC stars}. All are late WC, with
star K being the earliest WC star known in Wd~1 at WC7.}
\end{figure*}

Red spectra of WC stars are displayed in Fig.~4. The
lower-resolution spectrum of WR star N can be seen in
Fig.~3. Spectroscopic classification criteria  for WCL
stars between 6-10000{\AA} are discussed in \citetalias{one}. We
employ the  ratio between the Equivalent Width (EW) of \ion{C}{ii}~9900\AA\ 
and \ion{C}{iii}~9710\AA\ as a key diagnostic, found to be
$>0.14$ for WC9 stars.  

WR M was preliminarily classified WC9 by \citet{main} based on a
rather noisy $R$-band spectrum. Our new spectrum, shown in Fig.~4, for
which  EW(\ion{C}{ii})/EW(\ion{C}{iii}) $=0.17\pm0.03$, 
corroborates this classification, showing an obvious resemblance to
the spectra of the   
WC9 stars WR E and WR F displayed in \citetalias{one}. 

The new spectrum of WR C - which \citetalias{one} classified as WC8 -
 is  also presented 
 in Fig.~4. It  is clearly very similar to that of WR M, 
the main difference being the  slightly weaker
\ion{C}{ii} features. The EW(\ion{C}{ii})/EW(\ion{C}{iii}) ratio is
$0.12\pm0.03$, on the borderline between WC8 and WC9.

The new spectrum of WR K, also seen in Fig.~4, is clearly
earlier than those of WR M and WR C. The \ion{C}{ii} features are
clearly much weaker and the \ion{C}{iv}~7726\AA\  line is very
obviously present. The EW(\ion{C}{ii})/EW(\ion{C}{iii}) ratio is only
$0.03\pm0.01$, suggesting that WR K is WC7.

This is confirmed by the much higher resolution spectrum shown in
Fig.~5. The blend
around the position of H$\alpha$ is dominated by \ion{C}{ii}~6581\AA\
in WC9 spectra \citep{vreux83}, but in WR K, it peaks at
$\lambda$6568\AA\, as is typical of earlier-type WC stars, where it is
dominated by \ion{He}{ii}~6560\AA. The blend around $\lambda$7065\AA\
is likely dominated by \ion{C}{iv}~7062\AA.

Finally, we examine the newly discovered WR N, for which 
 we only have the  low-resolution grism\#1 spectrum shown in
Fig.~3. The ratio
between \ion{C}{iii}~5696\AA\ and \ion{C}{iv}~5812\AA\ is suggestive
of a WC9 spectral type. However, the absence of \ion{He}{i}~5875\AA\
supports an earlier type. As the S/N ratio is rather low in the
$\lambda\lambda5500-6000$\AA\ region, we prefer to use the red-end
classification criteria. The EW(\ion{C}{ii})/EW(\ion{C}{iii}) ratio is
$0.07\pm0.02$, typical of WC8 and indeed the strength of \ion{C}{ii}
features appears intermediate between that of WR M and WR K. We
therefore adopt WC8 for WR N.

\begin{figure}
\resizebox{\hsize}{!}{\includegraphics[bb=115 115 535 705, angle=-90]{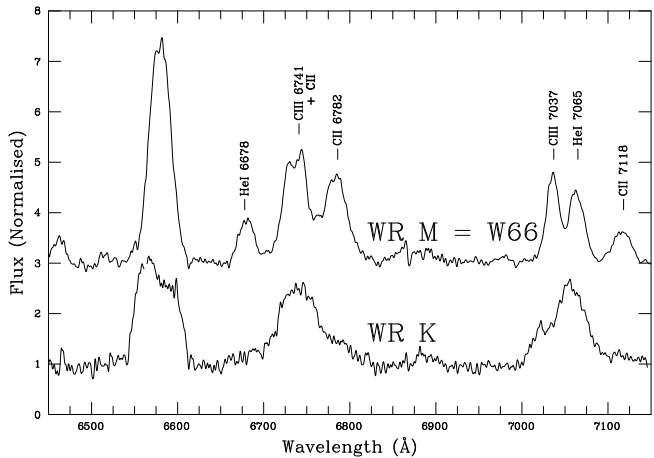}} 
\label{fig:highres} 
\caption{Intermediate resolution $R$-band spectrum of two WC stars in
Wd~1. WR M = W66 displays a typical WC9 spectrum. The spectrum of WR K
clearly implies an earlier spectral type. All the \ion{C}{ii} transitions are
very weak or absent and the emission lines are broad, several of them
dominated by \ion{C}{iii} and \ion{C}{iv} transitions (see text for details).}
\end{figure}

\subsubsection{WN stars}
\label{sec:wns}

Spectra of four  WN stars observed with grism \#4 are 
displayed in Fig.~4, while two other objects observed at
lower resolution are shown in Fig.~3. Of the four newly
discovered WN stars, the best 
S/N ratio has been achieved for WR Q, which we therefore choose to discuss 
first. We find its spectrum to be  extremely similar to
that of the WN6 star WR 85 displayed by \citet{vreux83}. Indeed the
weakness of \ion{He}{i}~7065\AA\ and \ion{He}{i}~6678\AA\ +
\ion{He}{i}~6683\AA\ argues against a spectral type later than
WN7. Unfortunately, the region covered by our spectra is not very
sensitive to the spectral type of WN stars and so a spectral type WN7
cannot be ruled out. 

The spectrum of WR star R (also in Fig.~4) has a much lower
S/N ratio, but does not appear to differ in any important respect from
that of WR star Q, leading to the same classification.

For WR star O, we have a low-resolution spectrum reaching
$\lambda$5500\AA\ (Fig.~3). 
The general aspect is very similar to those of WR
stars R and Q. Both  \ion{He}{i}~5875\AA\ and \ion{C}{iv}~5808\AA\
appear to be absent (within the -- rather large -- uncertainty allowed
by the limited S/N ratio), supporting a WN6 classification.

The spectrum of WR star P (also in Fig.~3) is very
different. The strength of all the \ion{He}{i} lines supports a WN8
spectral type. Specifically, only the WN8 stars in the catalogues of
\citet{vreux83} and \citet{vreux89} fulfil  the condition \ion{He}{i}~6678\AA
(+\ion{He}{ii}~6683\AA) $\la$ H$\alpha$ ((+\ion{He}{ii}), as happens
in WR star P.

WR star B was previously unclassifiable, due to the extremely low S/N 
ratio of the available spectrum. Our new spectrum, presented in 
Fig.~4, reveals  the  emission lines of this star to  be 
very weak, strongly suggesting the 
presence of an OB companion in the spectrum. Within the very limited
S/N ratio, \ion{He}{i}~7065\AA\ seems to be rather strong compared to 
\ion{N}{iv}~7114\AA, suggesting a spectral type WN8, but an exact
classification cannot be given.

Finally, we present a new higher S/N spectrum of WR star A,  the only 
broad-lined object in our sample
(Fig.~4). The spectrum does not offer any clues for
classification, except for the fact that it should be earlier than
WN7.

\subsubsection{The WNVL candidates}
\label{sec:wnvls}

Clark et al. (2004) identified the  first example of a WNVL star within 
Wd~1, the WN9 object W44 (= WR star L), while speculating that the 
emission line star W5 may possess an even later classification. New low 
resolution spectra (Table 1; not presented here) confirm the line 
identifications previously reported for both stars -- with the addition 
of weak He\,{\sc i} 7282{\AA} emission in W44 --  the emission lines 
appearing  to be narrow and single peaked.  In Fig.~6, we
present a  
higher resolution spectrum of W5 which reveals a wealth of new details, 
including P Cygni He\,{\sc i} emission lines and emission from low 
excitation metallic species such as N\,{\sc ii} and C\,{\sc ii}. In 
particular we  are able to  confirm the presence of the strong emission 
feature at $\sim$7235{\AA}, which  we  attribute to a C\,{\sc ii} doublet. 

\begin{figure} 
\resizebox{\hsize}{!}{\includegraphics[bb=55 80 500 760, angle=-90]{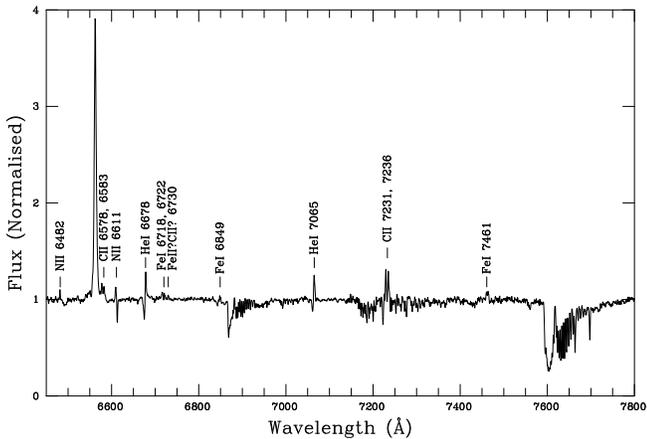}} 
\label{fig:w5} 
\caption{Intermediate resolution spectrum of the candidate WNVL star
  W5, showing tentative identifications for emission lines. Note the
  P-Cygni profiles in some of the \ion{He}{i} transitions.} 
\end{figure} 

As with W44, the lack of N\,{\sc iv} 7116{\AA}  precludes a classification 
earlier than WN9 for W5.  The H\,{\sc i} and He\,{\sc i} emission
spectrum is similar to  that of the WN11 star \object{S119}, although
the presence and  strength of the C\,{\sc  ii} emission
suggests a  temperature lower than the  
$24\:$kK inferred for \object{S119} (Paul Crowther, private 
communication 2004). If W5 is cooler than $24\:$kK, it is not
expected to display \ion{He}{ii} emission. Observationally, this is
difficult to check, as \ion{He}{ii}~4686\AA\
is outside our spectral range and \ion{He}{ii} lines in the red would
be expected to be very weak. A lack of
\ion{He}{ii} emission would prevent a WR classification, making W5 an
early  B supergiant. In this case, the strength of 
the  emission lines and the lack of a P-Cygni profile in H$\alpha$ would 
indicate a very extreme early B0--0.5Ia$^{+}$ classification (Paul Crowther, 
private communication 2004). Given this 
uncertainty, we choose to  denote  W5 as {\em Candidate} WR star S and 
adopt a provisional classification of WNVL/early BIa$^{+}$, apparently 
intermediate between the {\em bona fide} WN9 star  W44 and the 
extreme B5Ia$^{+}$ stars W7 and W33 \citep{main}.

\begin{table*}
\caption{Summary table for the currently identified Wolf-Rayet
  population within Westerlund 1; new or revised results from this
  paper are presented in {\bf bold} face.
  Columns 1 \& 2 present the various identifiers (including tentative numbers in the catalogue system of \citet{hucht}). Co-ordinates
  (J2000) are presented in Column~3. For objects correctly identified
  in \citet{main}, co-ordinates are determined  from 3.6-cm radio
  images (Dougherty, Clark \& 
  Waters; in prep.; formal errors are $\sigma_\alpha = \pm0.003^{\rm
  s}$, and $\sigma_\delta= \pm0.04^{\prime\prime}$). For new
  identifications, co-ordinates are simply extracted from the
  astrometric solution of the VLT/FORS2 image. Both solutions agree to
  better than $0\farcs2$. Co-ordinates for
  uncertain counterparts (as defined in Sect.~\ref{sec:res}) are given in  
  italics. Where available,
  broadband $VRI$ photometry is presented in Columns $4-6$. Finally the
  spectral classifications are presented  in Column~7.} 
\begin{center}
\begin{tabular}{lcccccccc}
\hline
WR star & Alternative &$\alpha$ & $\delta$  & V  & R  & I &Spectral \\
& Names&              &          &           &   &          & Type \\
\hline
A & WR\,77s, W72&16h47m08.32s & $-45\degr50\arcmin45\farcs5$ & 19.69 & 16.59 & 13.68 &{\bf $<$WN7} \\
B & WR\,77n &16h47m05.35s & $-45\degr51\arcmin05\farcs0$ & 20.99& 17.50 & 14.37 &{\bf WN8?} \\
C & WR\,77l &16h47m04.40s & $-45\degr51\arcmin03\farcs8$ & - &-  &- &  {\bf WC8.5} \\
D & WR\,77q &{\it 16h47m06.24s} & {\it $-$45$\degr$51$\arcmin$26$\farcs$5} &   -& - & - & WN6-8 \\
E & WR\,77p,W241 &16h47m06.06s & $-45\degr52\arcmin08\farcs3$  & -& -& -  & WC9\\
F & WR\,77m, W239&16h47m05.21s & $-45\degr52\arcmin25\farcs0$  & 17.86 & 15.39& 12.90 &  WC9 \\
G & WR\,77i &{\it 16h47m04.02s} & {\it $-$45$\degr$51$\arcmin$25$\farcs$2} & 20.87 & 17.75 & 14.68 & WN6-8  \\
H & WR\,77k &{\bf {\it 16h47m04.1s}} & {\bf  $-${\it 45}$\degr${\it 51}$\arcmin${\it 20}$\farcs${\it 0}} & - & - & - & WC9  \\
I & WR\,77e &{\it 16h47m01.67s} &{\it  $-$45$\degr$51$\arcmin$19$\farcs$9}  & -& -& - & WN6-8 \\
J & WR\,77c &{\it 16h47m00.89s} & {\it $-$45$\degr$51$\arcmin$20$\farcs$9}  &  -& - & - & WNL \\
K & WR\,77g &{\bf 16h47m03.1s} & {\boldmath $-45\degr50\arcmin43\arcsec$}  & -& -& -&  {\bf WC7}  \\
L & WR\,77j, W44 & 16h47m04.20s & $-45\degr51\arcmin07\farcs0$ & 18.86 & 15.61 & 12.52 &  {\bf WN9}  \\
M & WR\,77h, W66 & 16h47m04.0s & $-45\degr51\arcmin37\farcs5$ & 19.79 & 16.85& 13.96 &  {\bf WC9}  \\
{\bf N} &WR\,77b &{\bf 16h46m59.9s} & {\boldmath $-45\degr55\arcmin26\arcsec$} & - & {\bf 16.9}& {\bf 13.0} & {\bf WC8}   \\
{\bf O} &WR\,77r &{\bf 16h47m07.6s} & {\boldmath $-45\degr52\arcmin36\arcsec$} & -& -& -& {\bf WN6}  \\
{\bf P} & WR\,77d, W57c &{\bf 16h47m01.5s} & {\boldmath $-45\degr51\arcmin45\arcsec$}   & -& -& -& {\bf WN8} \\
{\bf Q} &WR\,77a &{\bf 16h46m55.4s}  & {\boldmath $-45\degr51\arcmin34\arcsec$}  & {\bf 20.3} & {\bf 17.5}& {\bf 14.7} & {\bf WN6-7}    \\
{\bf R} & WR\,77o, W14c &{\bf 16h47m06.0s} & {\boldmath $-45\degr50\arcmin22\arcsec$}  & -& -& -&  {\bf WN6-7}   \\
 {\it S} &   WR\,77f, W5 & 16h47m02.97s & $-45\degr50\arcmin19\farcs5$ & 17.49& 14.98 & 12.48 &  {\bf WNVL}  \\
\hline
\end{tabular}
\end{center}
\end{table*}

\section{Discussion \& Concluding Remarks}

Our current census for the WRs within Wd1  consists of 2 
transitional/WNVL(9-11), 9 WNL(6-8), one indeterminate WN (the 
broad-lined WR star A) and 7 WCL (7-9) stars. As such, we appear to be 
lacking early WR stars of both flavours. This may be most simply
explained by  the intrinsic  faintness of such objects. Given a median
value of $V-M_{V} \sim25.3$ for Wd~1 \citep{main}, assuming the
absolute visual magnitudes for early WR stars presented by
\citet{hucht} yields  
apparent $V$-band magnitudes $\ga22$, clearly beyond the reach of our
current photometry and spectroscopy. Theoretical models \citep{mm03}
also predict that WNE stars will
be rare compared to both WNL and WCL  stars,  due to their
shorter lifetimes.

With the above considerations in mind, it is of interest to compare
 the currently identified 
 WR population of Wd~1 to those of other young massive clusters in the
 Galaxy. First, we 
 would like to note that the resolution of W14 into three
 components, including a WNL, and the consequent refutation of a possible
 early O supergiant classification removes the sole
 observational datum that might suggest non-coevality for Wd~1. As
 discussed in \citet{main}, an age of 3.5-5~Myrs may safely be 
 estimated for Wd~1. Therefore we  
expect its population to differ from those of the younger NGC\,3603 and 
Arches clusters, which are found to be dominated by WN stars
 \citep[e.g.,][]{crowther,figer02}. 
A more revealing comparison may be established with the Quintuplet 
cluster \citep{figer} and the concentration of massive stars 
within the central parsec of the Galactic Centre
\citep[e.g.,][]{horrobin,genzel}, which have estimated ages similar to
those of Wd~1.

 Excluding the five Quintuplet Proper 
Members, which may be exceptionally dusty WCLs, the Quintuplet hosts 6 
WN and 5 WCL stars \citep{figer,homeier}. While the
population of WC stars in both the  
the Quintuplet and Wd~1 consists exclusively of WC7-9 stars, five of
the six Quintuplet WN stars are classified as WN9 or later, with only one 
earlier WN6 object (\citealt{figer}, and refs. therein); by  
comparison,
we have only found two $\geq$WN9 stars within Wd~1, with 9 of the 
remaining 10 objects being WN6-8. 

Inevitably, this comparison is prone to  concerns due to
completeness and selection effects. Given that the majority 
of the currently identified non-dusty WRs are amongst the faintest
spectroscopically surveyed stars within the Quintuplet, it is possible
that a further population of earlier WN stars
lies below current detection thresholds. Moreover, \citet{figer}
infer very high intrinsic luminosities for the WNL 
stars within the Quintuplet when compared to those within Wd~1.
Further
observational and analytic efforts  -- to arrive at a common detection
threshold and to determine if
the stellar parameters of the WNLs in both clusters systematically differ
--  are required before meaningful conclusions as to the relative WR
populations of both clusters may be drawn.

In contrast, the evolved  population of the central parsec -- as currently 
determined -- is remarkably similar to that of Wd~1. To date, it
consists of 6 Ofpe/LBV\footnote{Compared to the WNVL/early BIa$^{+}$
  star W5, the LBV W243, the B5Ia$^{+}$ stars W7, 33 \& 42 and
  arguably the six YHGs within Wd~1 -- 
\citet{smith} claim that YHGs occupy a closely related {\em 
evolutionary}  state to LBVs.}, 8 WNL (7-9), 4 WNE (5/6), 10 WCL 
(8/9)\footnote{Additional dusty  objects such as IRS13 E3A \& B may also 
host WCL stars \citep{maillard}.} and 1 WCE 
(5/6) stars, of which  only the latter spectral type is not represented 
in Wd~1  (\citealt{genzel}; Genzel, priv. comm. 2004; \citealt{maillard}). 

In spite of this similarity, the Galactic Centre population may not
represent a good analogue for Wd~1, as the origin of these massive
evolved stars in the central parsec  
is at present unclear, with \citet{genzel} finding that 
they occupy two thin coeval discs, each with similarly-sized populations 
of evolved stars, about half the size of the currently identified 
population of Wd~1. One possibility for the formation of such a 
distribution is the spiral in of two clusters -- indeed a possible 
remnant of such a  disrupted cluster, IRS13, is found to be associated 
with one ring (Genzel, priv comm. 2004;
\citealt{maillard}). Alternatively, \citet{genzel} suggest that a
collision between two interstellar clouds  
resulted in  two counter-rotating discs of gas which then formed the 
(evolved) massive stellar population. In any event, the stars found within 
the central parsec do not appear to form a stellar cluster in the same manner
as apparently `monolithic' clusters such as Wd~1 and the Arches. 

We therefore find that, at present, Wd~1 possesses the largest WR 
population of any known galactic cluster. Indeed, given the 
combination of distance and high reddening to Wd~1, the intrinsic 
faintness of even WNL and WCL objects \citep{hucht}, crowding in the
central regions of the cluster and the somewhat {\em ad hoc} nature of
the current observations, we  
suggest that our current census likely remains significantly
incomplete. Further investigation is needed in order to achieve a full
characterisation of the WR population of Wd~1, but the results presented
here already show that this cluster can provide us with a unique
laboratory to study the evolutionary paths followed by massive stars. 
 
\begin{acknowledgements} 
 IN  is a researcher of the
programme {\em Ram\'on y Cajal}, funded by the Spanish Ministerio de
Educaci\'on y Ciencia  and the University of Alicante, with partial
support from the Generalitat Valenciana and the European Regional
Development Fund (ERDF/FEDER). 
This research is partially supported by the Spanish MEC under grant
AYA2002-00814. 

We are very grateful to Dr. Amparo Marco for her help during this
 observing run. The NTT team have provided excellent support for this
 project since its start. We would like to specifically thank John
 Willis and Emanuella Pompei for their dedicated support. We also
 thank Paul Crowther for many informative discussions on the
 nature and classification of WR stars, and Karel van der Hucht for comments on the manuscript . 
\end{acknowledgements}

\end{document}